\documentstyle[prl,aps,epsf]{revtex}
\begin{document}
\draft
\twocolumn[\hsize\textwidth\columnwidth\hsize\csname @twocolumnfalse\endcsname
\title{Experimental Test of the Inter-Layer Pairing Models for High-T$_c$
Superconductivity Using Grazing Incidence Infrared Reflectometry.}

\author{J. Sch\"utzmann,$^1$ H. S. Somal,$^1$ A. A. Tsvetkov,$^{1,2}$ 
D. van der Marel,$^1$ G. E. J. Koops,$^1$ N. Koleshnikov,$^3$ Z. F. Ren,$^4$ 
J.H. Wang,$^4$ E. Br\"uck,$^5$ and A. A. Menovsky$^5$}

\address
{$^1$ Materials Science Centre,Laboratory of Solid State Physics,
University of Groningen,\\
Nijenborgh 4, 9747 AG Groningen, The Netherlands.\\
$^2$P.N. Lebedev Physical Institute, Russian Academy of 
Sciences, Moscow, 117924 Russia\\
$^3$Inst. of Solid State Physics, Russian Academy of Sciences, Chernogolovka,
142432 Russia\\
$^4$Dept. of Chemistry, Suny at Buffalo, Buffalo NY 14260-3000, USA\\
$^5$Van der Waals-Zeeman Institute, University of Amsterdam, The Netherlands.}
\date{19 august 1996, Phys. Rev. B, to be published in 1997}
\maketitle
\begin{abstract}
From measurements of the far-infrared reflectivity at grazing angles 
of incidence with p-polarized light we determined the c-axis Josephson
plasma frequencies of the single layer high T$_c$ cuprates
Tl$_2$Ba$_2$CuO$_6$ and La$_{2-x}$Sr$_x$CuO$_4$.
We detected a strong plasma resonance at 50 cm$^{-1}$ for 
La$_{2-x}$Sr$_x$CuO$_4$ in excellent agreement with previously published
results. For Tl$_2$Ba$_2$CuO$_6$ we were able to determine an upper limit
of the unscreened c-axis Josephson plasma frequency 100 cm$^{-1}$ or a
c-axis penetration depth $> 15 \mu$m. The small value of $\omega_J$ 
stands in contrast to recent a prediction based on the inter-layer 
tunneling mechanism of superconductivity.
\end{abstract}
% Superconductivity/optical properties
% Optical properties of condensed matter
% Exch, corr., dielectric and magn functions, plasmons
% superconductivity/response to EM fields
%
%
\vskip2pc]
\narrowtext
A striking feature of the high T$_c$ cuprates is the strong anisotropy of the 
conductivity. Both the low conductivity in the c-direction (below the Mott 
limit) and the frequency dependence (no Drude peak) support the 'confinement' 
hypothesis\cite{chakra} stating that hopping of single electrons between 
neighboring CuO$_2$ planes is inhibited in the normal state due to spin-charge 
separation. As tunneling of pairs of carriers is still allowed, 
the formation of Cooper-pairs results in a gain of kinetic energy, 
which stabilizes the superconducting state. Therefore within the
Chakravarty-Anderson interlayer model the main contribution to the condensation
energy ($E_{cond}$) is due to the Josephson coupling between adjacent 
CuO$_2$ planes. Recently Anderson\cite{phil} pointed out that, since the 
Josephson coupling energy is proportional to the plasma frequency 
of the condensate ($\omega_J$), a proportionality should exist 
between $\omega_J$ and T$_c$ for the cuprates with one CuO$_2$ layer 
per unit cell. One of 
the best compounds for testing the confinement hypothesis appears to
be Tl$_2$Ba$_2$CuO$_y$ at optimal oxygen concentration due to its
high value of T$_c$ (85 K) and the large separation between CuO$_2$ planes
(11.57$\AA$). According to the ILT hypothesis \cite{phil,houston,leggett}:
 \begin{equation}
  \hbar^2 \omega_J^2 =  \eta E_{cond} \frac{16\pi de^2}{a^2}   
  \label{ilt}
 \end{equation}
where $\eta$ represents the fraction of the condensation energy contributed
by the ILT mechanism\cite{leggett}, $d$ is the spacing between adjacent 
CuO$_2$-planes and $a$ is the in-plane lattice parameter. The condensation 
energy ($E_{cond} \propto T_c^2$) can be obtained 
directly from the specific heat\cite{loram}, resulting in $E_{cond}=80\mu$eV 
per unit of CuO$_2$ for Tl$_2$Ba$_2$CuO$_y$ ($T_c$=85 K) and $13 \mu$eV 
for slightly underdoped La$_{2-x}$Sr$_x$CuO$_4$ ($T_c$=32 K). 
The above expression for $\omega_J$ is radically different from 
conventional BCS-theory, where, adopting a 'dirty-limit' scenario 
for the c-axis conductivity, $\omega_J^2$ follows from the Glover-Tinkham-Ferrel
sumrule\cite{glover} 
 \begin{equation}
  \hbar^2 \omega_J^2 = 4\pi^2 \hbar \sigma_n\Delta(0) 
  \label{bcs}
 \end{equation}
Using the ILT prediction (Eq.\ref{ilt}) we calculate that 
$\omega_J \approx 1500 \mbox{cm}^{-1}$ for
Tl$_2$Ba$_2$CuO$_y$ and 530 cm$^{-1}$ for 
La$_{2-x}$Sr$_x$CuO$_4$ (T$_c$=32 K). In contrast, using Eq.\ref{bcs} we
calculate $\omega_J \approx$ 200 cm$^{-1}$ and 230 cm$^{-1}$  
for Tl$_2$Ba$_2$CuO$_y$ and La$_{2-x}$Sr$_x$CuO$_4$ respectively. 
These values for $\omega_J$ should still be complemented with the atomic 
polarizabilities and optical-phonon parameters for each of these compounds
to obtain the full c-axis dielectric function
 \begin{equation}
  \epsilon_{c}=\epsilon_{\infty}-\frac{\omega_J^2}{\omega^2}+
               \frac{4\pi i}{\omega}\sigma_e+
     \sum_j \frac{S_j\omega_j^2}{\omega_j^2-\omega(\omega+i/\tau_j)}
  \label{eps}
 \end{equation}
from which follows the position of the screened Josephson plasma-resonance.
\\
Most experimental techniques aimed at measuring the c-axis penetration depth
or $\omega_J$ require single crystals of sufficient thickness (several mm) 
in the c-direction. Lacking large crystals of Tl$_2$Ba$_2$CuO$_6$ we adopted
a different approach: By measuring the intensities of p-polarized light 
reflected from the ab-surface at a grazing angle of incidence (80$^\circ$)  
we can probe the longitudinal optical modes for the electric field
vector along the c-axis including the c-axis Josephson plasmon. 
In this study we used several flux-grown Tl$_2$Ba$_2$CuO$_6$ 
single crystals from the same batch\cite{kolesh}. The 
typical size of the crystals is 2$\times$2 mm in the ab-plane and 0.1 mm 
along the c-direction. The field-cooled and zero-field cooled DC 
magnetic susceptibility (inset of Fig. 1) shows a 
10 K wide transition with an onset at 90 K. 
For the DC transport measurements four gold contacts were 
evaporated, followed by a mild baking at 250 $^o$C. The arrangement 
of the four voltage and current contacts was selected for obtaining 
an accurate value of $\rho_c$, but only relative changes of $\rho_{ab}$ 
could be obtained. Both $\rho_c$ and $\rho_{ab}$ 
have a linear temperature dependence (Fig. 1) as has been reported before for
this compound by a number of groups\cite{tlrho}. The drop in resistivity 
takes place between 95 and 85 K. Details about the 
preparation and characterization of the Tl$_2$Ba$_2$CuO$_6$ films 
(T$_c$ = 80 K) and of the  La$_{2-x}$Sr$_x$CuO$_4$ crystals were described in 
Refs.\cite{ren} and \cite{somal} respectively. 
\\
For a strongly anisotropic material with a high conductivity in the 
ab-plane and a low conductivity along the c-axis, minima are found 
in the reflectivity 
 \begin{equation}
  R_p = \left|\frac{Z_0-Z_s^p}{Z_0+Z_s^p}\right|^2
  \label{fresnel}
 \end{equation} 
when Re${Z_s^p}$ has a maximum. Here  $Z_0$ is the vacuum impedance. 
The surface impedance for p-polarized light at oblique incidence is 
 \begin{equation}
  Z_s^p = \frac{Z_0}{n_{ab}\cos\theta}
  \sqrt{1-\frac{\sin^2\theta}{\epsilon_c}}
  \label{pseudo}
 \end{equation}
where $\theta$ is the angle of incidence with the surface normal, and 
$n_{ab}^2 = \epsilon_{ab}$ the ab-plane component of the dielectric tensor.
The c-axis longitudinal modes appear as sharp resonances in the surface
resistance\cite{tsui}, which is proportional to the pseudo-loss function
 \begin {equation}
  L(\omega) =\mbox{Im}  
  e^{i\phi}\frac{
  \sqrt{1-\frac{\sin^2\theta}{\epsilon_c}}}
  {1+|Z_s^p/Z_0|^2}
  = \frac{(1-R_p)|n_{ab}|\cos\theta}{2(1+R_p)}
  \label{pseudo2}
 \end{equation}
$L(\omega)$ is roughly the same as the c-axis loss function Im$(-1/\epsilon_c)$
apart from a weakly frequency dependent phase-shift
$\phi=\pi/2-\mbox{Arg}(n_{ab})$. The second equality in Eq.\ref{pseudo2}
can be used to calculate $L(\omega)$ from the grazing incidence
reflectivity data. Although this requires an estimate of $|n_{ab}|$, 
as $|n_{ab}|$ has a smooth frequency-dependence, the
multiplication with this factor does not influence the
position and line shape of the c-axis longitudinal modes.
\\
The p-polarized reflectivity was measured at an angle of incidence of 
$\theta = 80^\circ$. To increase the probing area, we mounted a mozaic
of four crystals which was fixed on top of a cone in order to reject
stray light. Absolute reflectivities were obtained by referencing the 
intensity reflected from the samples to the reflectivity after Au-coating 
the samples {\em in situ}. 
%The {\em in-situ} Au evaporation was performed 
%without rotating or translating the sample-rod between the sample and
%reference measurements. This procedure to a large extent compensates 
%diffraction effects, typically seen below 150 cm$^{-1}$ for samples with a 
%diameter of 2 mm. 
%
In Fig. 2 we present the grazing incendence reflectivity of the 
Tl$_2$Ba$_2$CuO$_{6+\delta}$ crystal. R$_p$ is characterized by a strong 
increase with decreasing temperature, 
which is caused by the temperature dependence of 
the in-plane optical conductivity. For $T = 6 K$ we observe 
deep and narrow minima\cite{phonons} at 157, 429, 631 cm$^{-1}$, which 
lie close to the out-of plane frequencies (143, 451 and 648 cm$^{-1}$) 
obtained from lattice dynamical calculations\cite{kulkarni}. 
Surprisingly an LO apex-oxygen bending mode predicted at 348 cm$^{-1}$ with 
a relatively large oscillator strength is absent in our data, 
similar to what was observed for Tl$_2$Ba$_2$Ca$_2$Cu$_3$O$_{10}$\cite{tl2223}. 
From Eq.\ref{eps} we see, that the half-width of the LO-phonons
in the loss function (Im$1/\epsilon_c$) is given by
$1/\tau+4\pi\sigma_e \epsilon_{\infty}^{-1} S/(\epsilon_{\infty}+ S)$, 
where $\tau$ is the intrinsic phonon life-time, $S$ the oscillator 
strength, and $\sigma_e$ the electronic optical conductivity. Hence
the width of these peaks can be used to estimate $\sigma_e$. After
correcting for peak-asymmetries of the pseudo-loss function
introduced through the phase-shift $\phi$, we obtained 
that $\sigma_e=0.7 \pm 0.3$ S/cm near $500\mbox{cm}^{-1}$,
{\em i.e.} below the DC conductivity at 100 K (2 S/cm). 
Two strongly damped modes at 86 and 538 cm$^{-1}$ are close to 
the calculated in-plane mode frequencies (84 and 560 cm$^{-1}$).
Having established these best-fit
parameters for the optical phonons we were able to determine the electronic
contribution of the in-plane optical conductivity by fitting the 
experimental data over a wide frequency range (50 to 6000 cm$^{-1}$) to 
Eq.\ref{fresnel}. Our estimated in-plane optical conductivity is in 
excellent agreement with recent results by Puchkov {\em et al.} 
\cite{puchkov}.     
\\
Although in the normal state the free carrier c-axis plasmon is
known to be overdamped in the cuprates, in the superconducting state 
we expect to observe a plasma minimum in R$_p$  
at the screened plasma resonance $\omega_{ps}/\sqrt{\epsilon_S}$\cite{tamasaku}
where $\epsilon_S$ is the background dielectric function (see Eq.\ref{eps}).
To demonstrate the reliability of the grazing incidence method we
display in the lower part of Fig. 3 R$_p$ measured on the 
$ab$-plane of a La$_{2-x}$Sr$_{x}$CuO$_4$ 
single crystal. Below T$_c$ a minimum occurs, which sharpens 
and shifts to 50 cm$^{-1}$ upon reducing temperature, 
consistent with the temperature dependence of $\omega_J$
found from normal incidence reflectivity measurements\cite{kim} 
on the ac-plane of this crystal. Such a structure is not present in the 
far-infrared or mid-infrared  
reflectivity spectra for the Tl$_2$Ba$_2$CuO$_6$ single crystal 
(Fig. 2) in the measured frequency range. Since at low frequencies
the signal/noise ratio is limited by the small size of the crystals, 
we also measured R$_p$ of thin films of Tl$_2$Ba$_2$CuO$_6$ in the
frequency range from 20 to 8000 cm$^{-1}$. We observed the same overall
behaviour and optical phonon spectrum as for the single crystal 
except for an additional optical phonon of the SrTiO$_3$-substrate  
at 170 cm$^{-1}$. The reflectivity is shown on an expanded scale in the
upper panel of Fig. 3. Even with the strongly improved signal/noise
ratio no clear evidence for a Josephson plasmon is present in these
data, except perhaps for a rather broad and shallow mimimum at 40 cm$^{-1}$.
\\
To enable a direct comparison to theory, we display in
Fig. 4a the c-axis pseudo loss function for Tl$_2$Ba$_2$CuO$_6$, 
both in the normal and in the superconducting state, calculated from the 
reflectivity spectra (Fig. 2) using Eq.\ref{pseudo2}, while adopting a smoothly
frequency dependent $ab$-plane dielectric function $\epsilon_{ab}$
fitted to $R_p$ over a wide frequency range upto 8000 cm$^{-1}$.
In Figs. 4b-4d we display model calculations of the pseudo-loss function based 
on the two scenario's (BCS and ILT) outlined in the introduction. We used
the optical phonon parameters discussed above\cite{phonons}. The phase
shift $\phi$ was calculated from the in-plane dielectric function
mentioned above.
In Figs. 4c and 4d the superconducting response was calculated
with a Josephson plasma frequency calculated from the BCS expression 
(Eq.\ref{bcs}). In Fig. 4c it was assumed that there is no residual 
c-axis optical conductivity at the position of the Josephson plasma 
resonance (50 cm$^{-1}$). In 4d a small residual conductivity of 3 S/cm was 
added near the resonance. Such a residual conductivity is expected in the 
case of d-wave pairing in the presence of impurity scattering. 
We notice that the occurrance of the Josephson plasmon at 50 cm$^{-1}$ has no
noticable effect on the position of the three longitudinal optical
phonons at 155, 430 and 630 cm$^{-1}$. 
The solid curve in Fig. 4b represents a simulation of the superconducting state
adopting $\omega_J = 1500$ cm$^{-1}$ implied by the ILT model (Eq.\ref{ilt}), 
while keeping all other c-axis parameters unaltered. The three peaks are now
of mixed phonon-plasmon character. The large value of the Josephson plasma 
frequency results in a considerable shift of the two peaks at 430 and 630 
cm$^{-1}$ towards lower frequencies, and a strong suppression of intensity 
of the 155 cm$^{-1}$ mode. The peak at 700 cm$^{-1}$ is essentially 
a screened superfluid plasmon. 
\\
Experimentally no Josephson plasmon
is observable in Fig. 2, 4a or for the film in Fig. 3 except perhaps
for a rather broad feature around 40 cm$^{-1}$ in Fig. 3. Taken together
with the fact that there is no observable shift of the LO phonons at 
146, 379, and 594 cm$^{-1}$ as the temperature is reduced from 300 K to 
6 K, an upper limit can be set for the screened plasma resonance 
of 40 cm$^{-1}$. Using the fitted phonon-parameters and $\epsilon_{\infty}$
for $\vec{E}\parallel \vec{c}$\cite{phonons}, this implies that the
unscreened Josephson plasma frequency satisfies $\omega_J < 100 cm^{-1}$, 
or equivalently the c-axis penetration depth satisfies 
$\lambda_c > 15 \mu$m for $T\rightarrow 0$. 
For La$_{2-x}$Sr$_x$CuO$_4$ and Tl$_2$Ba$_2$CuO$_6$ 
the fraction of the condensation energy contributed by the ILT mechanism 
$\eta =(\omega_J^{exp}/\omega_J^{ILT})^2\approx 0.2$ and 
$\eta < 0.005$ respectively. Tl$_2$Ba$_2$CuO$_6$ differs from 
La$_{2-x}$Sr$_x$CuO$_4$ in other respects: The very low 
normal state $c$-axis conductivity ($< 2 S/cm$ at 100 K) combined with
a linear rise as a function of temperature has been noticed before and
is confirmed both from our DC measurements and from the dynamical conductivity
around 500 cm$^{-1}$ obtained from our analyzes of the line shape
of the prominent longitudinal optical phonons. These observations,
and the very low c-axis Josephson plasma frequency definitely pose a
challenge both to 'conventional' mechanisms and to the implementation 
of the ILT model as it was described in Ref.\cite{chakra}. 
To comply with these facts within the context of an ILT mechanism
one may postulate\cite{philpriv} 
the existance of small electron pockets in the TlO layers,
which are also two-dimensional and cross the main CuO band only at a few 
points in 2D $k$-space. The matrix element which would uncross these bands
is then restricted in 2D $k$-space, which leads to a reduced value of 
$\omega_J^2$, while  T$_c$ depends mainly on those parts of 
Fermi surface where the crossing occurs. Within this scenario  
large portions of the Fermi surface should exist 
where $|\Delta| \ll |\Delta_{max}|$, which can be tested experimentally. 
\\
We measured the c-axis infrared properties of thin plate-like
Tl$_2$Ba$_2$CuO$_6$ single crystals and thin films and the ab-face of
La$_{2-x}$Sr$_x$CuO$_4$ using grazing incidence infrared spectroscopy. 
The screened plasma resonance of Tl$_2$Ba$_2$CuO$_6$ was found to be below 
40 cm$^{-1}$, corresponding to an unscreened c-axis
Josephson plasma frequency of 100 cm$^{-1}$ and a lower limit of the
c-axis penetration depth of 15 $\mu$m.  
These experimental results, together with the low value of the conductivity
along the c-direction, pose a challenge both to conventional theories and
the inter layer tunneling mechanism.
\\
{\em Acknowledgements}
One of us (D.v.d.M) gratefully acknowledges numerous fruitfull discussions 
with P.W. Anderson and A.J. Leggett at various stages of this project.
This investigation was supported by the Netherlands Foundation for
Fundamental Research on Matter (FOM) with financial aid from
the Nederlandse Organisatie voor Wetenschappelijk Onderzoek (NWO) and 
by the EEC (Human Capital and Mobility). The work performed at SUNY/Buffalo 
was supported by NYSERDA,ORNL,ANL,ONR,and NSF.
\begin{figure}[th]
    \begin{center} 
     \leavevmode
     \epsfxsize=85mm
     \epsfysize=65mm
%     \hbox{\epsfbox{tlfig1.ps}}
    \end{center} 
 \caption{dc resistivity and dc susceptibiltiy measured with H$\|$ab (10 Oe)
  of a Tl$_2$Ba$_2$CuO$_6$ single crystal.}
 \label{fig1}
\end{figure}
\begin{figure}[th]
    \begin{center} 
     \leavevmode
     \epsfxsize=85mm
     \epsfysize=70mm
%     \hbox{\epsfbox{tlfig2.ps}}
    \end{center} 
 \caption{Reflectivity R$_p$ of a Tl$_2$Ba$_2$CuO$_6$ single crystal 
  measured with p-polarized light incident on the ab-surface at an
  angle of 80$^\circ$.}
 \label{fig2}
\end{figure}
\begin{figure}[th]
    \begin{center} 
     \leavevmode
     \epsfxsize=85mm
     \epsfysize=82mm
%     \hbox{\epsfbox{tlfig3.ps}}
    \end{center} 
 \caption{Reflectivity R$_p$ of a Tl$_2$Ba$_2$CuO$_6$ 
  thin film (upper panel) and of a La$_{1.85}$Sr$_{0.15}$CuO$_4$
  single crystal measured with p-polarized light incident on the 
  ab-surface at an angle of 80$^\circ$. }
 \label{fig3}
\end{figure}
\begin{figure}
   \begin{center} 
     \leavevmode
     \epsfxsize=85mm
     \epsfysize=104mm
%     \hbox{\epsfbox{tlfig4.ps}}
    \end{center} 
 \caption{Pseudo loss function $L(\omega)$ of  Tl$_2$Ba$_2$CuO$_6$: 
  (a) experimental data at 100 K (dashed curve) and 6 K (solid), 
  (b) simulation for the normal state (dashed) and for the
  superconducting state (solid) adopting the ILT value for 
  $\omega_J$ of 1500 cm$^{-1}$, 
  (c) simulation of the superconducting state adopting the BCS value for
  $\omega_J$ of 200 cm$^{-1}$,
  (d) {\em idem.} with a residual Drude conductivity of 3 S/cm added close to
  the plasma resonance. In (b),(c) and (d) $\sigma_e$ was set to 0.5 S/cm near 
  the dominant phonon frequencies.}
  \label{fig4}
\end{figure}

\begin{references}
%
\bibitem{chakra}
 S. Chakravarty and P.W. Anderson, 
 Phys. Rev. Lett. {\bf 72}, 3859 (1994).
%
\bibitem{phil} 
 P.W. Anderson, 
 Science {\bf 268}, 1154 (1995).
%
\bibitem{houston} 
 D. van der Marel, 
 {\em 10th Anniversary HTS Workshop on Physics, Materials 
 and Applications}, Houston, March 1996, in press. 
%
\bibitem{leggett}
 A.J. Leggett, 
 submitted to Science (1996).
%
\bibitem{loram}  
 J. Loram, private communication.
%
\bibitem{glover}  
 M. Tinkham, {\em Introduction to Superconductivity} 
 (Mc Graw-Hill, New York, 1996)
%
\bibitem{kolesh}
 N. N. Koleshnikov {\em et al.}, 
 Physica C {\bf 242} 385 (1995).
%
\bibitem{tlrho}
 H.M. Duan {\em et al.}, Physica C {\bf 185}, 1283 (1990);
 T. Manako {\em et al.}, Physica C {\bf 185}, 1327 (1990). 
%
\bibitem{ren}
 C. A. Wang {\em et al.}, Physica C {\bf 262}, 98 (1996).
%
\bibitem{somal} 
 H.S. Somal {\em et al},
 Phys. Rev. Lett. {\bf 76} 29 february, (1996).
%
\bibitem{tsui} 
 O.K.C. Tsui, N.P. Ong, and J.B. Peterson,
 Phys. Rev. Lett. {\bf 76}, 819 (1996).
%
\bibitem{phonons} 
 Using Eq.\ref{eps} the fit parameters of the phonons are 
 $\epsilon_{\infty} = 4.0$ and
 $\{\omega_{Tj}(\mbox{cm}^{-1}),S_j,1/\tau_j(\mbox{cm}^{-1})\} =   
 \{146,0.30,1\},\{379,0.93,8\},\{594,0.31,8\}$ $(j=1..3)$.
%
\bibitem{kulkarni} 
 A.D. Kulkarni {\em et al},
 Phys. Rev. B {\bf 41}, 6409 (1990).
%
\bibitem{tl2223} 
 Jae Kim {\em et al},
 Phys. Rev. B, 49 (1994) 13065-13069
%
\bibitem{puchkov} 
 A.V. Puchkov {\em et al}, 
 Phys. Rev. B {\bf 51}, 3312 (1990).
%
\bibitem{tamasaku} 
 K. Tamasaku, Y. Nakamura, and S. Uchida, 
 Phys. Rev. Lett. {\bf 69} (1992) 1455
%
\bibitem{kim} J. H. Kim, A. Wittlin, D. van der Marel {\em et al.},
 Physica C {\bf 247}, 297 (1995).
%
\bibitem{philpriv}
 P.W. Anderson, private communication.
\end{references}
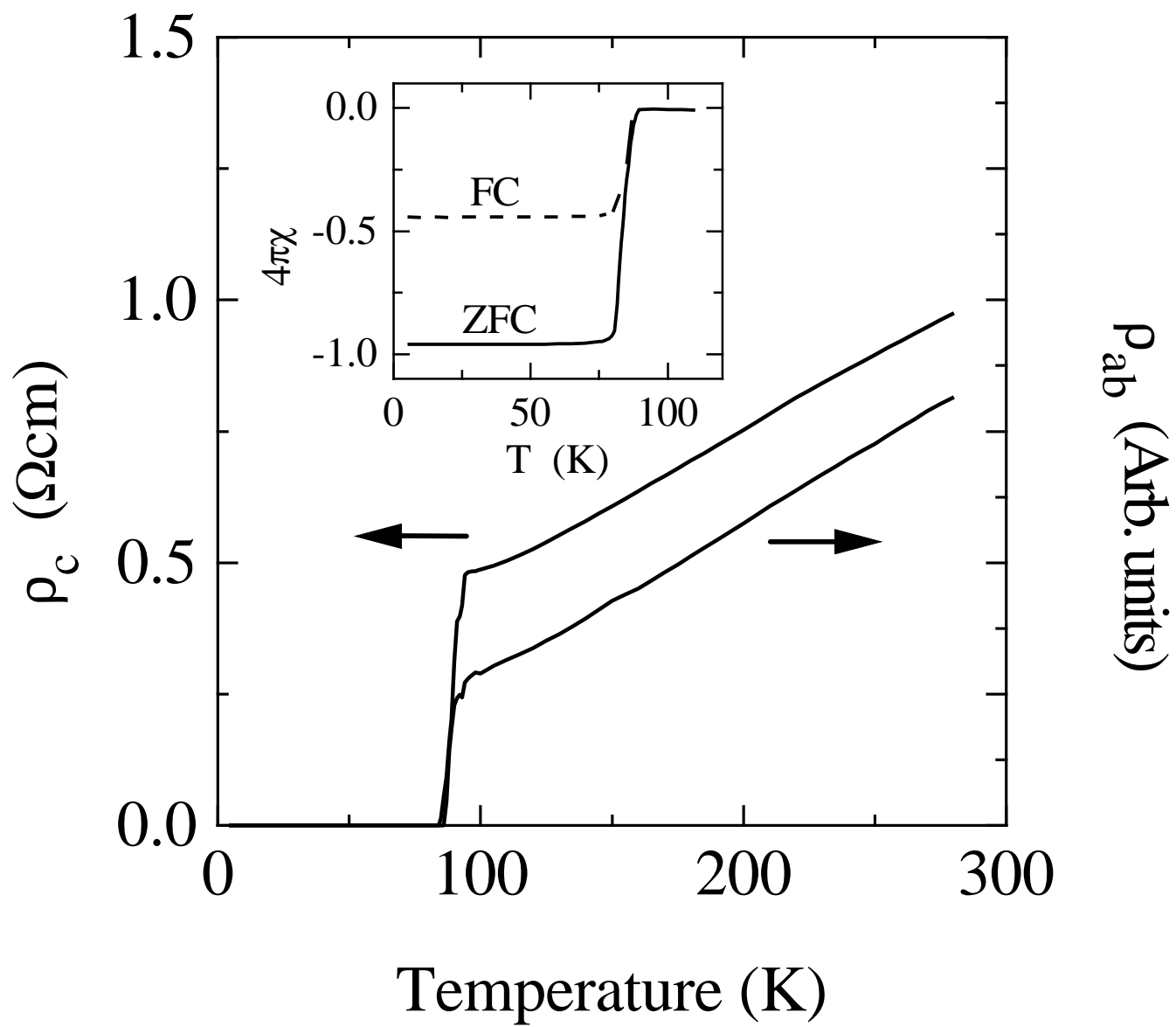
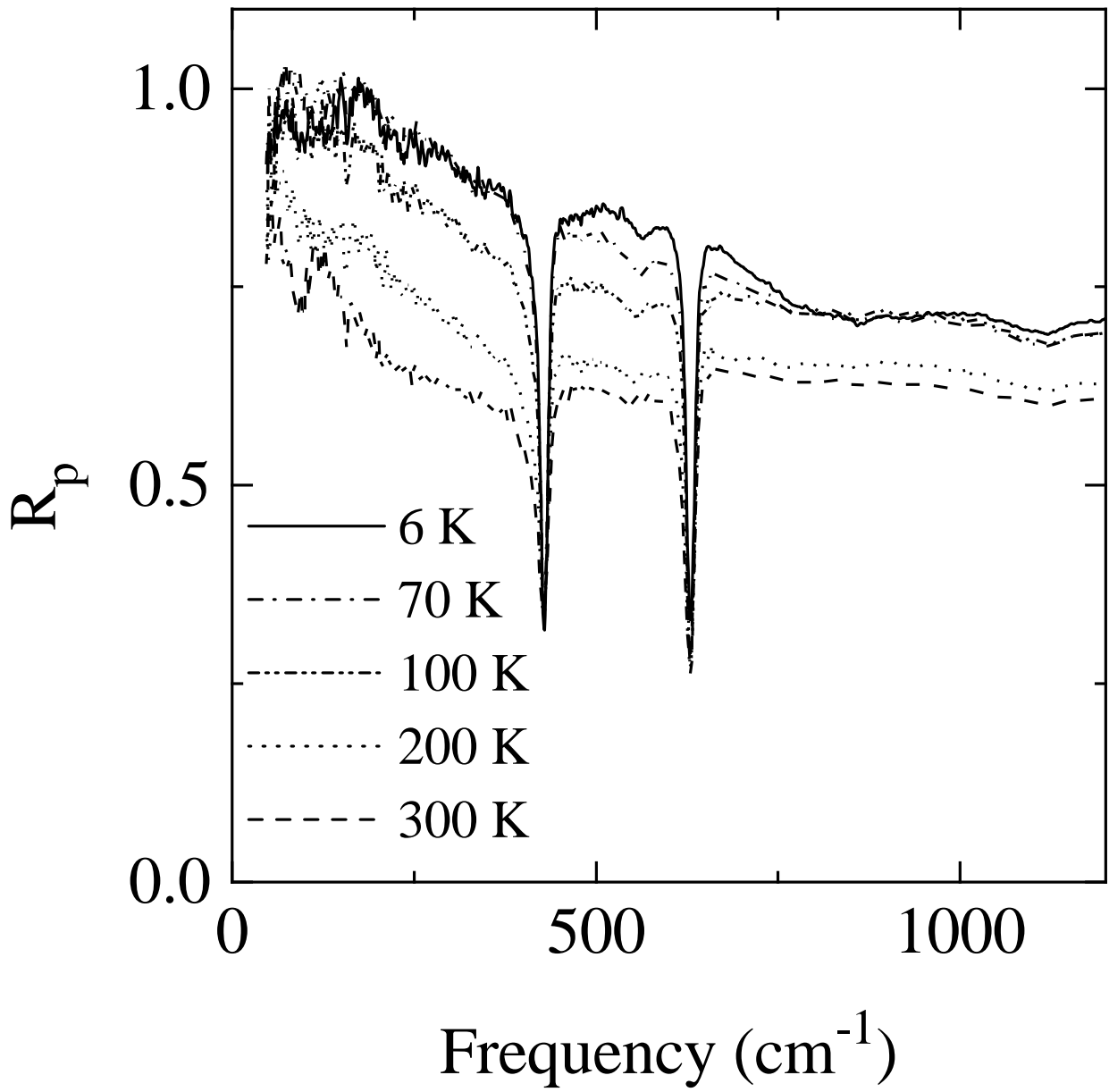
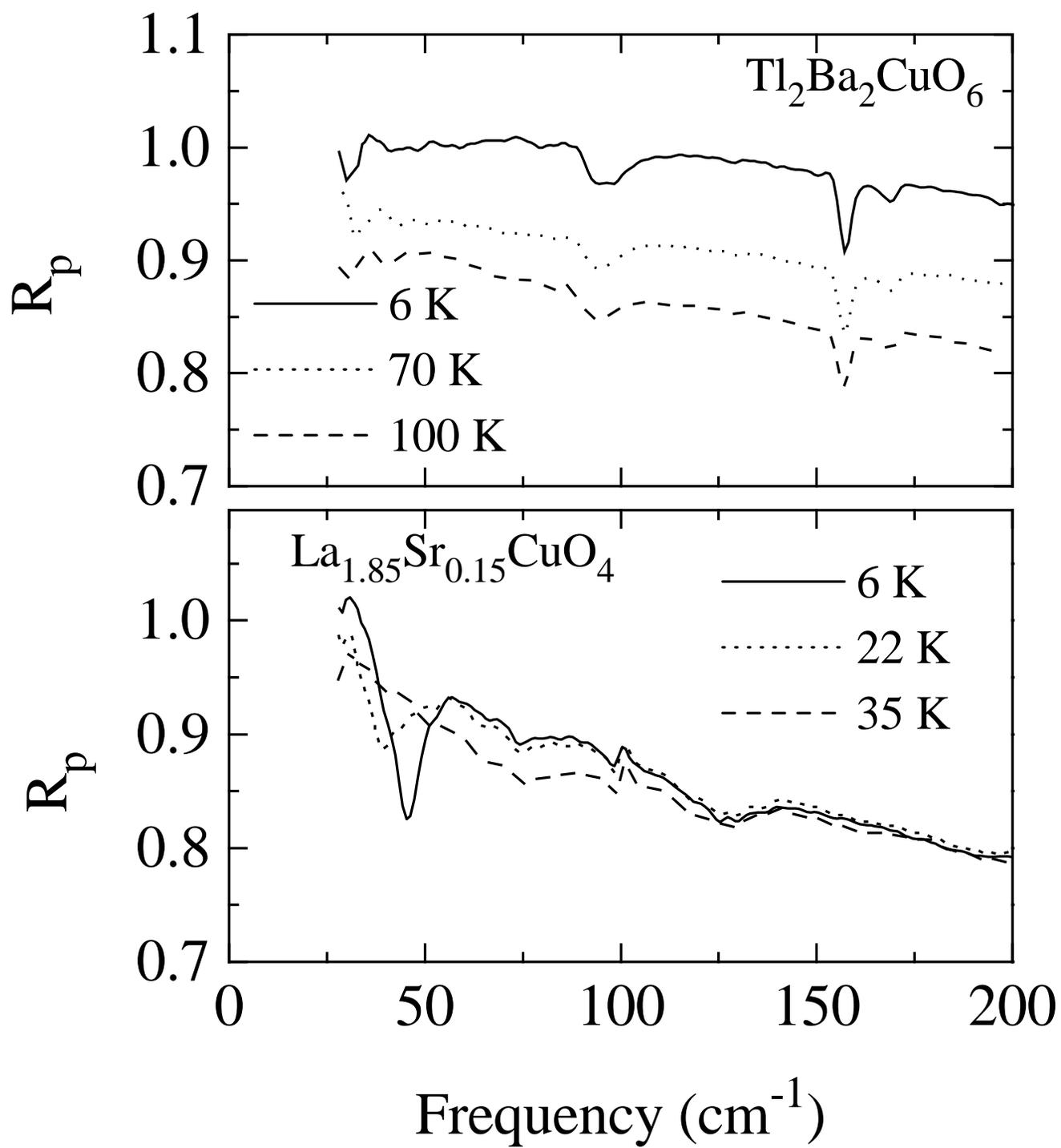
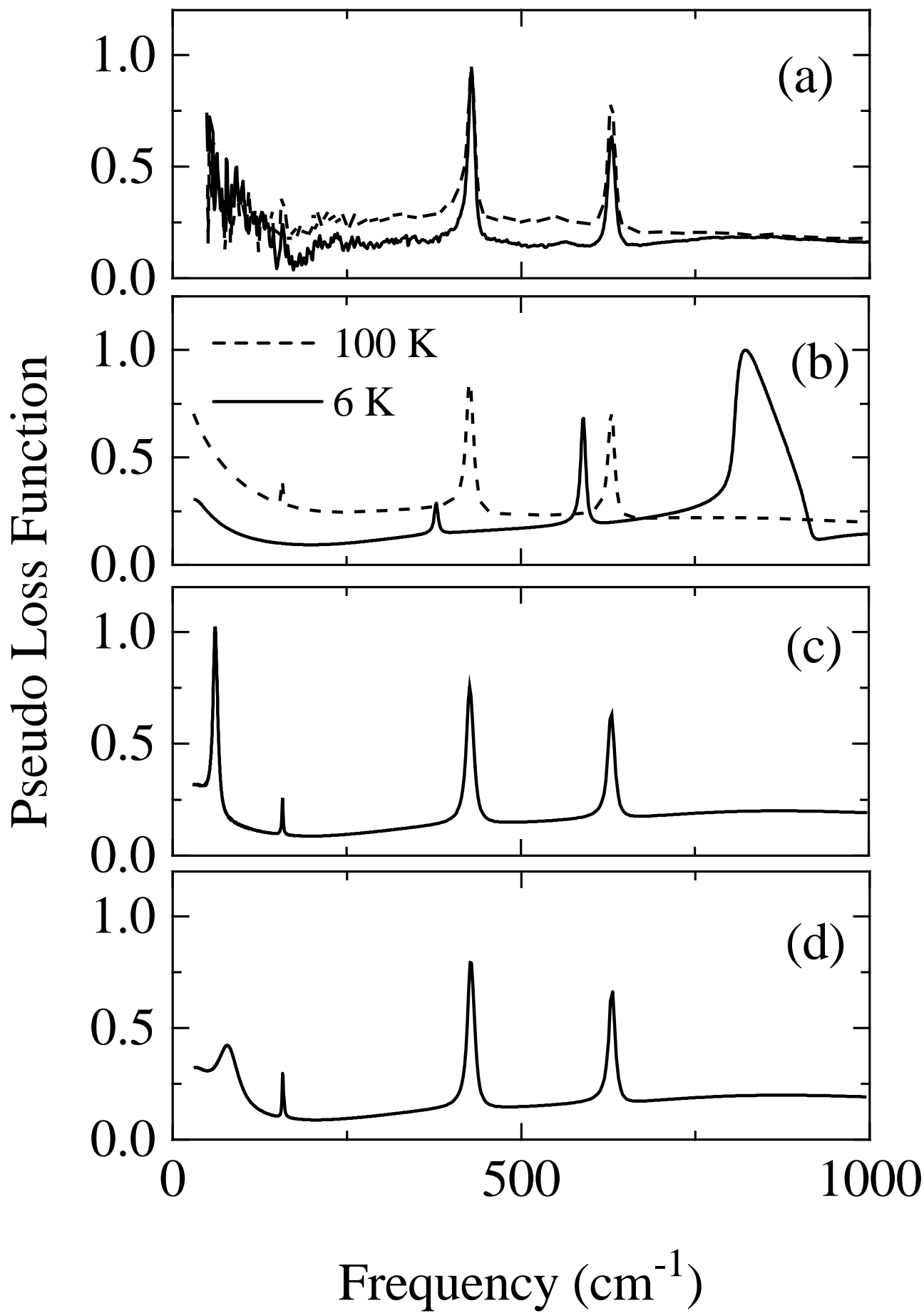
\end{document}